\documentclass[preprint2]{aastex631}

\let\tablenum\relax
\usepackage{siunitx}
\usepackage{placeins}
\usepackage{mathtools}
\shorttitle{Theoretical half-life for $^{134}$Cs and $^{135}$Cs in astrophysical scenarios}
\shortauthors{Taioli et al.}
\begin{document}

\title{Theoretical estimate of the half-life for the radioactive $^{134}$Cs and $^{135}$Cs in astrophysical scenarios}

\author[0000-0000-0000-0000]{Simone Taioli}
\affiliation{European Centre for Theoretical Studies in Nuclear Physics and Related Areas (ECT*), Bruno Kessler Foundation}
\affiliation{Trento Institute for Fundamental Physics and Applications (TIFPA-INFN), Trento, Italy} 
\email{taioli@ectstar.eu}

\author[0000-0003-0309-4666]{Diego Vescovi}
\affiliation{Goethe University Frankfurt, Max-von-Laue-Strasse 1, Frankfurt am Main D-60438, Germany}

\author[0000-0001-8944-5820]{Maurizio Busso}
\affiliation{Department of Physics and Geology, University of Perugia, Via A. Pascoli snc, I-06123 Perugia, Italy}
\affiliation{INFN, section of Perugia, Via A. Pascoli snc, I-06123 Perugia, Italy}

\author[0000-0001-5386-8389]{Sara Palmerini}
\affiliation{Department of Physics and Geology, University of Perugia, Via A. Pascoli snc, I-06123 Perugia, Italy}
\affiliation{INFN, section of Perugia, Via A. Pascoli snc, I-06123 Perugia, Italy}

\author[0000-0001-9683-9406]{Sergio Cristallo}
\affiliation{INAF, Osservatorio Astronomico d’Abruzzo, Via Mentore Maggini snc, I-64100 Collurania, Teramo, Italy}
\affiliation{INFN, section of Perugia, Via A. Pascoli snc, I-06123 Perugia, Italy}

\author[0000-0002-2537-0038]{Alberto Mengoni}
\affiliation{ENEA, Dipartimento Fusione e Tecnologie per la Sicurezza Nucleare, Via Martiri di Monte Sole 4, I-40129 Bologna, Italy}
\affiliation{INFN, section of Bologna, Viale Berti Pichat 6/2, I-40127 Bologna, Italy}

\author[0000-0001-9287-8651]{Stefano Simonucci}
\affiliation{School of Science and Technology, University of Camerino, Via Madonna delle Carceri 9B, I-62032 Camerino, Macerata, Italy}
\affiliation{INFN, section of Perugia, Via A. Pascoli snc, I-06123 Perugia, Italy}
\email{stefano.simonucci@unicam.it}

%% Mark off the abstract in the ``abstract'' environment. 
%\linenumbers
\begin{abstract}

We analyze the  $^{134}_{55}$Cs$\rightarrow^{134}_{56}$Ba and $^{135}_{55}$Cs$\rightarrow^{135}_{56}$Ba $\beta^-$ decays, which are crucial production channels for Ba isotopes in Asymptotic Giant Branch (AGB) stars.  We reckon, from relativistic quantum mechanis, the effects of multichannel scattering onto weak decays, including nuclear and electronic excited states (ES) populated above $\simeq$ 10 keV, for both parent and daughter nuclei. 
We find increases in the half-lives for $T>10^8$ K (by more than a factor 3 for $^{134}$Cs) as compared to previous works based on systematics. We also discuss our method in view of these previous calculations.
An important impact on half-lives comes from nuclear ES decays, while including electronic temperatures yields further increases of about 20\% at energies 10-30 keV, typical of AGB stars of moderate mass ($M \lesssim 8~M_{\odot}$). Despite properly considering these effects, the new rates remain sensitively lower than the TY values, implying longer half-lives at least above 8-9 keV.
Our rate predictions are in substantial accord with recent results based on the shell model, and strongly modify branching ratios along the $s$-process path previously adopted. With our new rate, nucleosynthesis models well account for the isotopic admixtures of Ba in presolar SiC grains and in the Sun. 

\end{abstract}

%% Keywords should appear after the \end{abstract} command. 
%% The AAS Journals now uses Unified Astronomy Thesaurus concepts:
%% https://astrothesaurus.org
%% You will be asked to selected these concepts during the submission process
%% but this old "keyword" functionality is maintained in case authors want
%% to include these concepts in their preprints.

%% From the front matter, we move on to the body of the paper.
%% Sections are demarcated by \section and \subsection, respectively.
%% Observe the use of the LaTeX \label
%% command after the \subsection to give a symbolic KEY to the
%% subsection for cross-referencing in a \ref command.
%% You can use LaTeX's \ref and \label commands to keep track of
%% cross-references to sections, equations, tables, and figures.
%% That way, if you change the order of any elements, LaTeX will
%% automatically renumber them.
%%
%% We recommend that authors also use the natbib \citep
%% and \citet commands to identify citations.  The citations are
%% tied to the reference list via symbolic KEYs. The KEY corresponds
%% to the KEY in the \bibitem in the reference list below. 

\section{Introduction} \label{sec:intro}
New scenarios to unravel heavy-element nucleosynthesis in stars were recently opened by extended surveys of spectroscopic observations \citep{2020ApJS2493A}, as well as by the analysis of presolar grains formed in stellar winds, which were trapped in pristine meteorites \citep{ours}, offering new constraints on the isotopic abundances generated in stellar processes \citep{busso+21}.

After decades of experiments on reaction-rates for neutron-captures and charged-particle processes, our understanding of the above scenario now depends also on improving the knowledge of weak nuclear reactions in hot stellar plasmas. Their accurate assessment still represents a crucial bottleneck, while experiments simulating stellar conditions in terrestrial ionized plasmas to measure decay rates are still in their infancy \citep{PANDORA,mascali22}. In particular, for a better agreement between nucleosynthesis models and observations \citep{vescovi2019effects} a great deal of theoretical work to assess stellar reaction rates with increased accuracy is still required \citep{chemev}; in fact, there is a dearth of computations for decay processes in stellar scenarios from first-principles \citep{simonucci2013theoretical,vescovi2019effects}. 

In this respect, we notice that $\beta$-decay spectra of allowed and forbidden transitions are typically calculated by using an analytical expression of the rate \citep{RevModPhys.90.015008}, where semi-empirical factors account for the nuclear structure, the phase-space, and the atomic exchange. However, not even this simpler approach is implemented in stellar neutron capture computations, where, instead, the phenomenological compilation \citep{TAKAHASHI1983578, TAKAHASHI1987375} (hereafter TY) is universally used. There, unknown transition strengths are estimated by analogy to laboratory decays of nearby nuclei with similar transitions. In a recent study for example 
\citep{Li_2021}, new $\log(ft)$ values were calculated from the nuclear shell model to obtain $^{134}$Cs stellar rates within this standard approach \citep{RevModPhys.90.015008}. In this work the electronic degree of freedom (DOF) is neglected, despite its crucial contribution to the rate. Nevertheless, the authors already show significant improvement in the astrophysical application, with the rate changing in the same direction as found using our analysis. Recently developed approaches, based on large-scale diagonalization shell model calculations, can be also used \citep{Langanke_2021}, even though they have seen so far applications only to part of the nuclide chart. 

One element requiring such studies is certainly Barium. In AGB stars, its abundance depends solely on slow ($s$) process nucleosynthesis that after the $n$-capture on stable $^{133}$Cs meets a branching point at $^{134}$Cs, where further $n$-captures compete with $\beta$-decay, e.g. the
$_{55}^{134}\mathrm{Cs}(J^{\pi}=4^{+}) \longrightarrow _{56}^{134}\mathrm{Ba}(J^{\pi}=4^{+}) + e^- + \bar{\nu}$ allowed transition.
Another reaction branching occurs at $^{135}$Cs, $\beta$-decaying to $^{135}$Ba via a 2$^{\mathrm{nd}}$ forbidden unique transition Cs$(7/2^+) \longrightarrow$ Ba$(3/2^+)$. The resulting balance, which should yield production of 100\% of the $s$-only $^{134}$Ba and $^{136}$Ba isotopes, is thus very delicate.
In particular, the $^{134}$Cs $\longrightarrow ^{134}$Ba decay (half-life 2.0652 y) is characterized by a Q-value (2058.7 keV; \citealt{SONZOGNI20041}) large enough to accommodate Cs high-lying nuclear excited states (ESs), e.g. $5^+$ at 11.2442 keV and $3^+$ at 60 keV above the ground state (GS). The emergence of $^{134}$Cs short-lived nuclear ESs, possibly populated at high temperature, and the decay to $^{134}$Ba isomers (such as $4^+$, $3^+$) may dramatically affect this transition owing to their different forbiddenness.

When $n$-capture on $^{134}$Cs feeds the longer-lived $^{135}$Cs, also the latter can decay through the mentioned Cs$(7/2^+)\longrightarrow$ Ba$(3/2^+)$ path. Its half-life is $2.3\times10^6$ y (with a GS-GS $Q$-value of 268.7 keV; \citealt{SINGH2008517}). The inclusion of several ES of $^{135}$Cs (notably the $5/2^+$ states at 249.767 and 408.026 keV above the GS), together with the population of electronic excited levels, may significantly affect the total rate at high temperature also for this nucleus. 

Furthermore, despite these transitions being quark-level processes, they also strongly depend on extra-nuclear factors, such as temperature and electron density, which significantly vary in the layers of evolved stars outside the degenerate core. Indeed, these parameters may affect the ionization degree of the atomic systems where the decay occurs, which in turn modifies the discrete--to--continuum transition ratio \citep{simonucci2013theoretical, palmerini2016lithium, vescovi2019effects}.

In this work we lay the foundation of a fully-relativistic quantum-mechanical method for calculating the temperature and density dependence of the $\beta$-decay half-life of s-process branching point isotopes, by considering both the electronic and nuclear ES population dynamics, and apply it here for the first time to $^{134}$Cs and $^{135}$Cs.
We show that these DOFs act concurrently to strongly change the Cs isotopes half-lives, even by several orders of magnitude, with respect to laboratory conditions and previous phenomenological approaches based on systematics (see TY). %\citep{TAKAHASHI1983578,TAKAHASHI1987375}.
We apply this approach to revise recent models aiming to explain the peculiar ratios of Ba isotopes in presolar SiC grains and the s-process origin of the $^{134}$Ba and $^{136}$Ba isotopes in the solar system \citep{ours,busso+21}. 

\section{Calculation of $\beta$-decay rates of $^{134}$Cs and $^{135}$Cs}\label{sec:calculation}
Our method is rooted in the single-particle mean-field (MF) approximation to the internuclear and interelectronic interactions, but is systematically and straightforwardly improvable by embedding higher accuracy many-body methods for including dynamical correlation \citep{HjorthJensen:391525}.
\begin{figure*}[t!]
\centering
\resizebox{\hsize}{!}{\includegraphics{{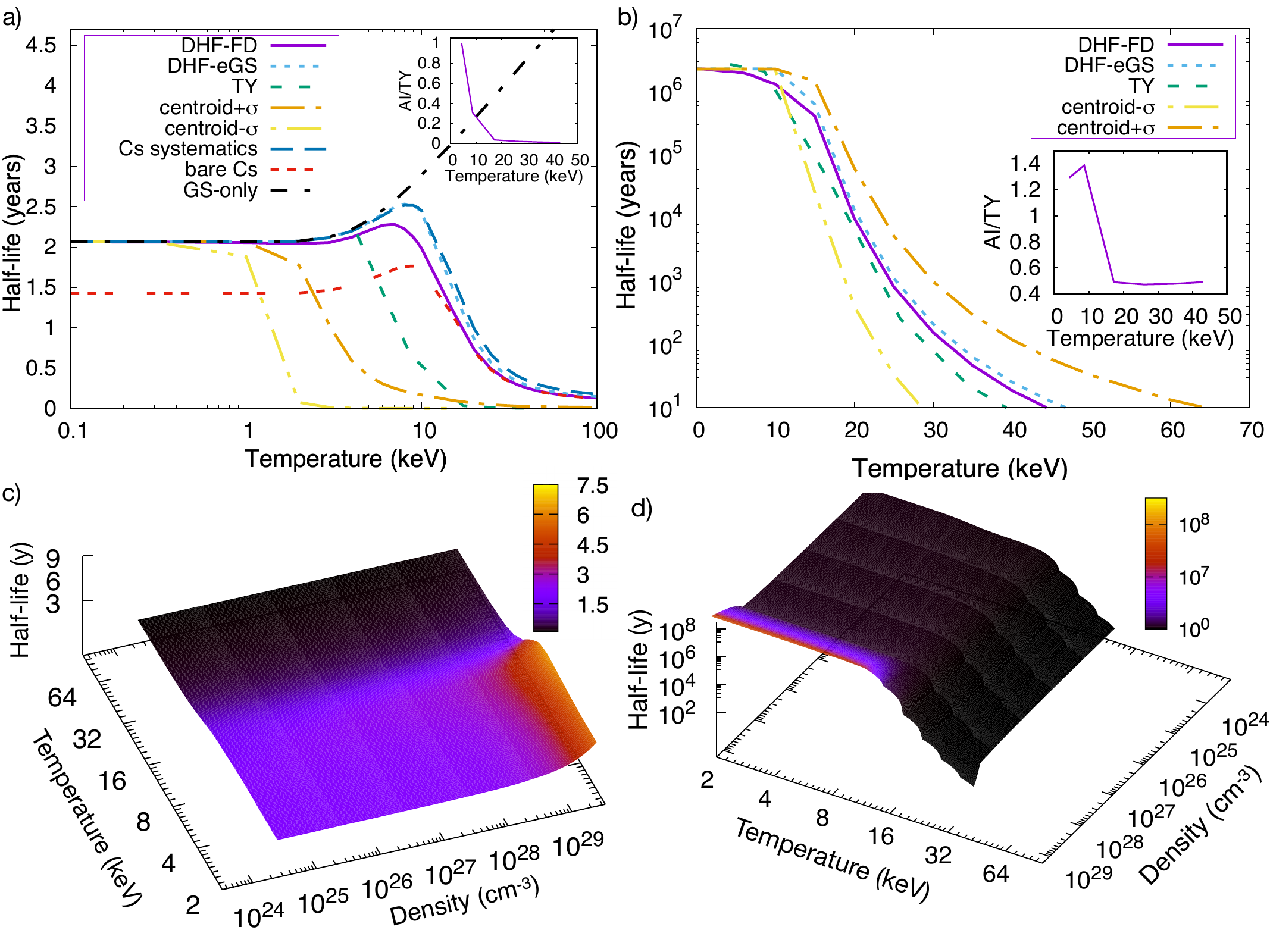}}}
%\plotone{Combined.pdf}
\caption{\label{fig1}
Half-life of $^{134}$Cs (a) and of $^{135}$Cs (b) vs. temperature (violet line) from our DHF calculations, for $n_p=10^{26}$ cm$^{-3}$. They include: (a) the nuclear states at $4^+,5^+,3^+$ of $^{134}$Cs and (b) those of $^{135}$Cs at $5/2^+,5/2^+$, as well as the electronic DOF. In cyan (DHF-eGS) we show the half-life computed when electrons are clamped in their GS. Orange and yellow lines show the maximum (centroid+$\sigma$) and minimum (centroid-$\sigma$) values obtained by taking the $\log(ft)$ from the general systematics, while in the blue line we used simply the centroid of the specific $^{134}$Cs systematics (this does not exist for $^{135}$Cs). Green lines show the previous estimates by TY.
%\citep{TAKAHASHI1983578,TAKAHASHI1987375}.
In panel a) we also show (red line) the half-life for a completely ionized $^{134}$Cs (bare Cs) and the nuclear GS to GS decay half-life (black line, GS-only). The trend is similar for $^{135}$Cs (not shown). Insets of a) and b) give the ratios between our ab-initio rates (AI) and TY %\citep{TAKAHASHI1983578,TAKAHASHI1987375} 
values for $n_p=10^{26}$ cm$^{-3}$. Panels c) and d) display $^{134}$Cs and $^{135}$Cs half-lives according to our model, including both nuclear and electronic DOFs, as a function of temperature and proton density.}
\end{figure*}  

Our approach, based on the first-principle calculation of the transition operator written in terms of the total QED Hamiltonian, proved accurate in reproducing a few terrestrial observables \citep{morresi2018nuclear}. Here is generalized to include the temperature and density dependence for dealing with astrophysical conditions (see Appendix~\ref{sec:app_theo} for further details).
Within our framework, the weak Hamiltonian is factorized into the product of leptonic and hadronic currents \citep{morresi2018nuclear}. We also require that the hadronic current is separable into neutron and proton field operators, which means essentially that the decaying neutron acts as a single-particle correlated to the {\it core} of the remaining nucleons only geometrically, so as to recover the experimental total angular momentum of the parent reactant. Furthermore, one can safely assume that the leptonic current can be written as a product of the neutrino and electron quantum field operators. The hadronic and leptonic currents are both reckoned by the self-consistent numerical solution of the Dirac--Hartree--Fock equation (DHF). 

Protons and neutrons are interacting via a semi-empirical relativistic Wood-Saxon potential, while the electron-electron Coulomb repulsion is modelled via a local density approximation to the electron gas ($V_{ex}\propto \rho (r) ^{1/3} $) \citep{Slater1951,Salvat1987}. Electrons populate the energy levels according to a Fermi-Dirac distribution, and the chemical potential is assessed by assuming that they behave as an ideal Fermi gas in a box with a relativistic energy-momentum dispersion. The non-orthogonality between the bound initial and final orbitals, the possible presence of shake-up (excitation) and shake-off (ejection) are also taken into account in the calculation of the continuum electron wavefunction. At high temperature, we considered positron formation, enforcing the overall neutrality of the plasma (see Table~\ref{tab:tableS1} reporting electron vs. proton densities at all temperatures).
For the $_{55}^{134}\mathrm{Cs} \longrightarrow _{56}^{134}\mathrm{Ba}$ transition, the lepton field carries no angular momentum. The rate has been assessed under the assumption, derived by the nuclear shell model, that the decaying neutron within the parent $^{134}$Cs occupies the $2d_{3/2}$ shell, while the generated proton populates the $1g_{7/2}$ orbital of $^{134}_{56}$Ba.
In principle, a superposition of single-particle near-by energy states mixed by the dynamical correlation, such as the $^{134}$Cs $1h_{11/2}$ and $3s_{1/2}$ shells, may contribute to the $\beta$ transition. However, owing to a bigger jump of $\Delta J$ the level of forbiddenness of the $\beta$ transition from those shells to the $^{134}$Ba $1g_{7/2}$ shell would be higher and, thus, the transition less likely to occur.

In our calculations we included the $4^+$ GS and the two lowest energy nuclear states ($5^+,3^+$) of $^{134}$Cs, occupied according to a Boltzmann distribution $f(E_i)=w_ie^{-E_i/K_{\text B} T}$, where $E_i$ is the energy of the $i$-th nuclear level, weighted by the relevant level degeneration factor ($w_i=9, 11, 7$, respectively for the 3 states mentioned above). 

To reproduce the experimental values at Earth conditions a constant correction factor of 4.4 to our calculated half-lives is applied at all temperatures. This renormalization factor is needed to correct the hadronic current, which may suffer from the MF evaluation of the nucleon wavefunctions. In our analysis we assume that the nuclear matrix elements do not depend on the temperature for a given transition because of the leptonic and hadronic currents factorization and of the independence of the single-particle orbitals of temperature. Indeed, the matrix elements of the hadronic currents only depend on the nuclear single-particle orbitals of the decaying neutron and of the proton before and after the decay, respectively. Once the initial parent and final daughter states are selected the hadronic current matrix elements are independent of temperature, and their energy scale (MeV) is rather different from those typical of AGB stars (keV). 
In light of these considerations in place, we can safely state that the correction factor is the same at all temperatures and the fact that terrestrial decays are determined by decay of atomic Cs does not play any role. The half-lives of $^{134}$Cs with and without the correction factors are reported in Tab. \ref{tab:table6} of the appendix B.

\begin{deluxetable}{ccc}[t!]
\tablecaption{Comparison between the $^{134}$Cs rates obtained by TY
%\citep{TAKAHASHI1987375}
and using our model (units in $s^{-1}\times 10^{-8}$) for different temperatures and fixed proton density $n_p=10^{26}$ cm$^{-3}$. Note that our rate remains always lower than TY suggest.
\label{tab:table1}}
\tablehead{
\colhead{$T_8$} & \colhead{TY} & \colhead{This work}}
\startdata
0.5~(4.31) & 1.02 &  1.02\\
1~(8.62) &  3.28 &  1.01 \\
2~(17.23) & 63.1  &  2.28 \\
3~(25.85) & 211.0  &  4.73 \\
4~(34.47)  & 481.0 &  7.22 \\
5~(43.09)  & 889.0 &  9.36 \\
\enddata
\tablecomments{$T_8$ is the temperature expressed in $10^8$ K, while the corresponding values in keV are in
parentheses.}
\end{deluxetable}
In Fig.~\ref{fig1}a and in the relevant inset as well as in Table~\ref{tab:table1} 
we compare our results with those
by TY.
A finer grid of
points can be found in Table~\ref{tab:table4} of the appendix B. We stress that in TY the effect
of the excited nuclear dynamics on the $\beta$ rates was estimated via the
$\log(ft)$ values for general $\beta$ transitions of given forbiddenness derived
from systematics. Moreover, at odds with our calculations, neither TY include
in their estimate the electronic DOF nor they calculate the nuclear matrix
elements from first-principles. 

We observe that the inclusion of the nuclear ES dynamics represents the most relevant effect on the $^{134}$Cs $\beta$-decay rate 
at high temperatures ($>$ 10 keV), where nuclear states are most 
likely populated. Indeed, the presence of fast-decaying nuclear ESs can increase the rate by a factor of 15 at 100 keV (1 GK) and up to 23 at 1000 keV, with respect to room temperature conditions. 
In particular, such half-life decrease can be almost entirely attributed to the $3^+$ nuclear excited state (ES), which implies a rate 80 times higher than the $4^+$ nuclear GS (if they were at each time the only occupied nuclear states to decay), while GS-only decay increases with temperature (as shown in Fig.~\ref{fig1}a). 

We also notice that the temperature has a relevant effect on the lepton DOF in the range [0-10] keV, as revealed in Fig.~\ref{fig1}a by the difference between the full half-life and the one with electrons in their GS. In the case of a bare nucleus, where all Cs electrons are stripped out (Cs 1s binding energy is ~36 keV), the half-life is consistently lower (by 20\% in laboratory conditions) than that one of the neutral atom, as the emitted $\beta$-electron can land in any bound orbital.
In Fig. \ref{fig1}a we also report the reference band of $^{134}$Cs half-life using the standard deviation (centroid $\pm \sigma$) of such systematics. We notice that TY recommendations are out of this range, while our results are in better agreement with the specific systematics of $^{134}$Cs (where the electronic DOF are of course neglected).

Finally, in Fig.~\ref{fig1}c we plot the $^{134}$Cs half-life vs. proton density and temperature as varying in the interior of stars. We notice that, at temperatures $>$ 10 keV, the half-life shows a significant drop, even at very high density, owing to the decay from the $^{134}$Cs nuclear ESs. 

\begin{deluxetable*}{@{\extracolsep{5pt}}ccccccccc}[t!]
\tablecaption{Comparison between the $^{135}$Cs rates obtained by TY %\citep{TAKAHASHI1987375} 
and using our model (units in $s^{-1}$) for different temperatures and proton densities. Both our rates are lower than those by TY at $T > 10^8$ K.\label{tab:table2}}
\tablehead{
\colhead{} & \multicolumn{2}{c}{$n_p=10^{26}$ cm$^{-3}$} & \multicolumn{2}{c}{$n_p=3\times 10^{26}$ cm$^{-3}$} & \multicolumn{2}{c}{$n_p=10^{27}$ cm$^{-3}$} & \multicolumn{2}{c}{$n_p=3\times 10^{27}$ cm$^{-3}$} \\
\cline{2-3} \cline{4-5} \cline{6-7} \cline{8-9}
\colhead{$T_8$} &
\colhead{TY} & \colhead{This work} &
\colhead{TY} & \colhead{This work} &
\colhead{TY} & \colhead{This work} &
\colhead{TY} & \colhead{This work} 
}
\startdata
0.5~(4.31) & 8.12e-15 & 1.05e-14 & 7.90e-15  & 1.02e-14 & 7.92e-15 & 9.70e-15 & 7.39e-15 & 9.11e-15\\
1~(8.62) & 1.04e-14  & 1.44e-14  & 8.78e-15 & 1.22e-14 &  8.04e-15 & 1.08e-14 & 7.81e-15 & 9.79e-15 \\
2~(17.23) &  6.91e-13 & 3.39e-13 & 6.65e-13 & 3.27e-13 & 6.09e-13 & 3.01e-13 & 5.52e-13 & 2.66e-13 \\
3~(25.85) & 8.64e-11 & 4.08e-11  & 8.55e-11 & 4.04e-11 & 8.24e-11 & 3.91e-11 & 7.74e-11 & 3.64e-11 \\
4~(34.47)  & 9.77e-10 & 4.66e-10 & 9.65e-10 & 4.64e-10 & 9.52e-10 & 4.57e-10 & 9.17e-10& 4.38e-10 \\
5~(43.09)  & 4.18e-09 & 2.05e-09  & 4.15e-09 & 2.05e-09 & 4.08e-09 & 2.03e-09 & 3.96e-09 & 1.97e-09\\
\enddata
\tablecomments{$T_8$ is the temperature expressed in $10^8$ K, while the corresponding values in keV are in
parentheses.}
\end{deluxetable*}

A similar analysis has been carried out also for the $^{135}$Cs $\beta$-decay. 
In Fig.~\ref{fig1}b we plot the half-life vs. temperature obtained by our DHF model as compared to TY estimates (color codes are explained in the caption). To reproduce the experimental values at Earth conditions a constant correction factor of $\approx$90 to our calculated half-lives is applied at all temperatures. The half-lives of $^{135}$Cs with and without the correction factors are reported in Tab. \ref{tab:table6} of the Appendix B. In this case, both data sets are found within the range spanned by the general systematics for these transitions. Again, we plot the half-life obtained with electrons clamped in their GS to determine the electronic contribution to the $\beta$ transition with increasing temperature. The latter is neglected in TY, while the inclusion of this DOF may almost halve the half-life around 10 keV. The results obtained with our model are compared with those by TY in Table~\ref{tab:table2} and in the inset of Fig.~\ref{fig1}b (a finer grid is reported in Table~\ref{tab:table5} of the Appendix B). 
In Fig. \ref{fig1}d) we report the half-life vs. temperature and proton density, finding similar, but steeper descent with increasing temperature with respect to $^{134}$Cs. \\

\section{Astrophysical implications and conclusions}\label{sec:conclu}

\begin{figure*}[t!]
\centering
\resizebox{\hsize}{!}{\includegraphics{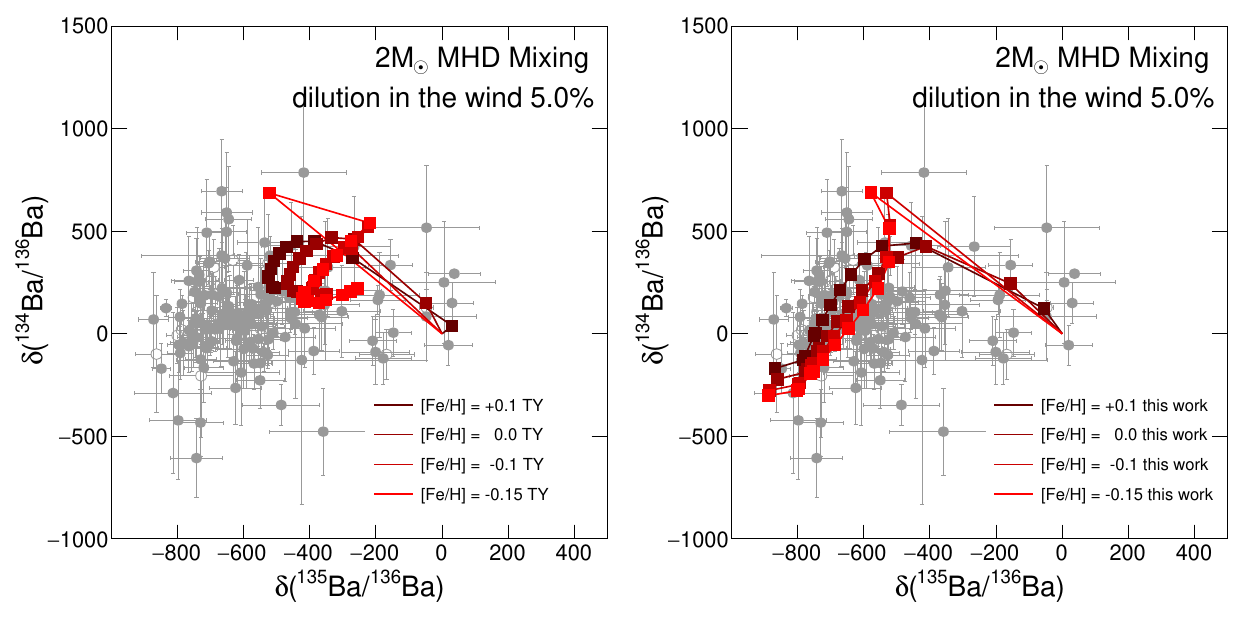}}
\caption{Left panel: isotopic ratios of $^{134}$Ba and $^{135}$Ba with respect to $^{136}$Ba, displayed as part-per-mil deviations from the solar
ratios (indicated by the symbol $\delta$) computed in recent AGB models \citep{ours}, with cross sections from the KADoNiS v1.0 database and decay rates from TY. Right panel: the results of the same models, where only the decay rates for $^{134}$Cs and $^{135}$Cs are changed, using those of the present work. We remind that computations in \cite{ours} are from 2 M$_{\odot}$ stars, where magneto-hydrodynamic processes induce the penetration of protons into He-rich layers, producing $^{13}$C then releasing neutrons through $^{13}\mathrm{C}(\alpha,n)^{16}\mathrm{O}$. Abundances are computed in stellar winds, where we assume that magnetic blobs further add 5\% of s-process
rich material in flare-like episodes. The symbol [Fe/H] indicates $\log(X_{\mathrm Fe}/X_{\mathrm H})_{star} - \log( X_{\mathrm Fe}/X_{\mathrm{H}})_{sun}$. Curves with different colors refer to stellar models of different metallicity.\label{grains}}
\end{figure*}

Our results immediately confirm recent suggestions \citep{busso+21}, according to which a more accurate treatment of $\beta$-decays in hot plasmas is needed for improving the modelling of Ba isotopes in AGB stars. This is especially true for the composition of presolar SiC grains \citep{ours}. We therefore used our rates to revise the calculations presented in \cite{ours}. 
In Fig.~\ref{grains} (left panel) we present the results obtained by using cross sections from the KADoNiS repository (Karlsruhe Astrophysical Database of Nucleosynthesis in Stars version v1.0, \citealt{cross}) and decay rates from TY. The right panel shows the changes obtained exclusively by adopting the  $\beta$-decay rates of this work. We stress that the new model curves, reproducing the expectations of AGB stars characterized by masses and metallicities as indicated, improve remarkably the agreement with observational constraints (gray dots) and fit well the main area occupied by the data. We notice that the novel rates presented here improve even on a tentative guess reported in \cite{ours} (see Fig. 19(a) therein). 

\begin{figure*}[t!]
%\centering
%\resizebox{\hsize}{!}{\includegraphics{solarSS1.pdf}}
\plotone{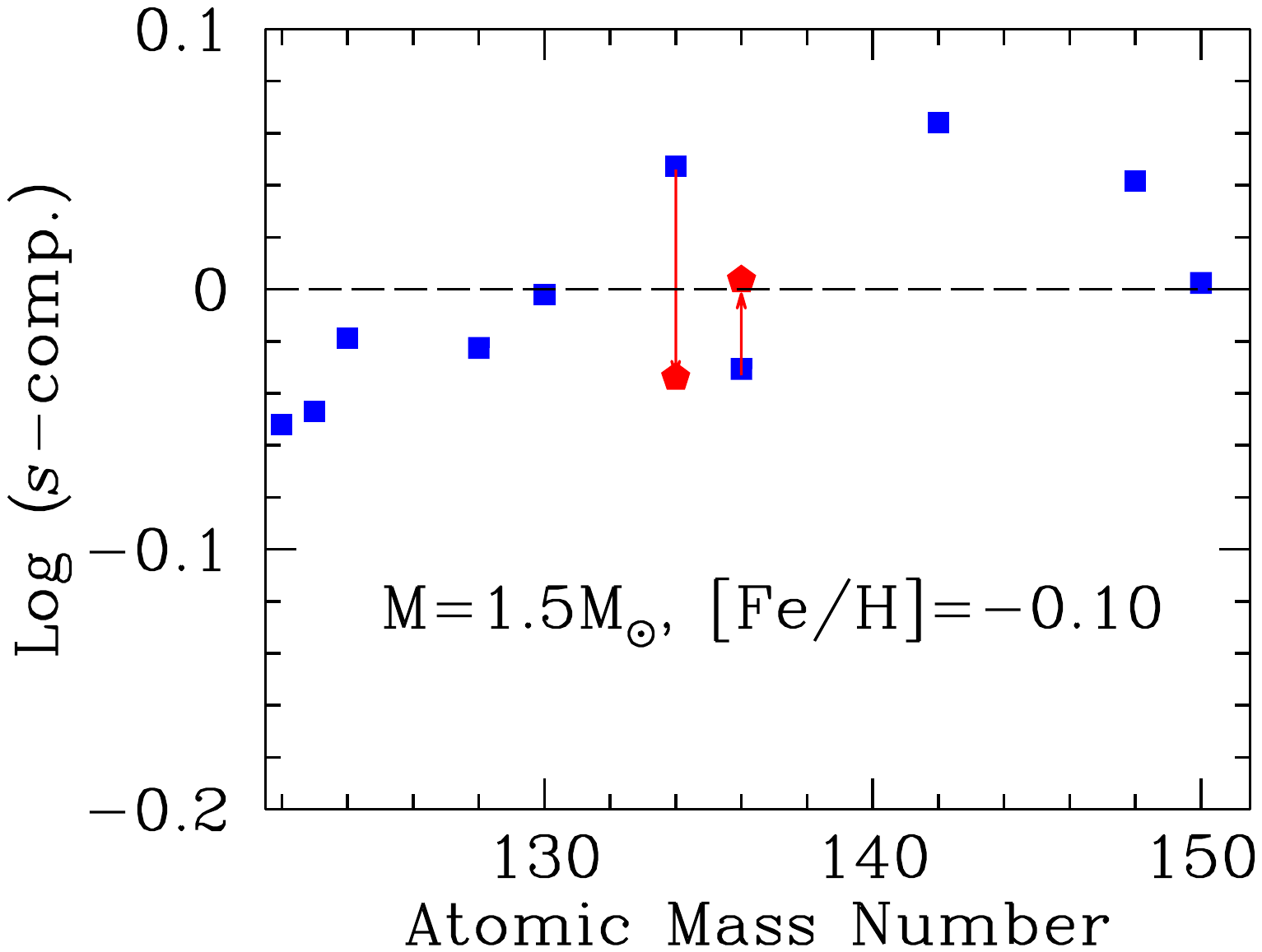}
\caption{\label{solar} Percentage of $s$-process contributions (blue dots) as computed by \cite{busso+21}, for $s$-only nuclei near the magic neutron number $N=82$. In a perfect model, without nuclear or stellar uncertainties, all points should be aligned at zero. Red points (marked by arrows) show the results for $^{134}$Ba and $^{136}$Ba in a revised calculation where, all the rest remaining the same, we adopted the $\beta^-$-decay rates for $^{134}$Cs and $^{135}$Cs computed in the present work. We note that two more nuclei look as outliers in the figure. They are $^{142}$Nd and $^{148}$Sm, whose abundances are critically linked to the decay rates of $^{142}$Pr, $^{147}$Nd and $^{148}$Pm, for which a revision like the present one would be needed.}
\end{figure*}

A further implication of our half-life revision concerns the solar distribution of $s$-elements. 
%Here, in particular, we perform some of the calculations carried out in \cite{busso+21,ours}.
%In particular, 
In this regard, very recently rather extensive analyses of neutron captures in AGB stars pointed out how a reappraisal of the rates here presented for $\beta^-$ decays, with a more accurate treatment of their dependence on temperature and density, was crucial for any improvement in our understanding of the isotopic admixture of Ba in the Sun \citep{busso+21}. In particular, it was shown that (see Fig. 4 in \citealt{busso+21}) an {\it average} model could be identified, suitable to mimic (if properly normalized) the outcomes of a simulation from a chemical 
evolution model of the Galaxy at least for the $s$-process contribution to heavy nuclei. 
In those estimates, $^{134}$Ba and $^{136}$Ba should be both at the level of one, as they do not receive contributions from other processes. 
%We notice that in Fig.~\ref{solar} the percentage of s-process contributions to the isotope production is plotted in $\log$ scale (thus for $\log(s)=0$ the contribution is s-only).
In Fig. \ref{solar} we show the results of the calculations carried out using the same physical models and nuclear parameters as in \cite{busso+21}, only updating the two rates discussed here. The percentage of s-process contributions to the isotope production is plotted in $\log$ scale (thus for $\log(s)=0$ the contribution is s-only). With blue dots we show the original results for $s$-only nuclei sited close to the {\it magic} neutron number $N=82$. In red we report the new values found for $^{134}$Ba and $^{136}$Ba using our new rates (all the rest remaining untouched). We stress that the improvement over the previous expectations is striking. We notice that similar improvements are also expected for Galactic Chemical Evolution models computed with full stellar evolutionary yields \citep{prantzos2020}, which were showed to have severe problems with $^{134}$Ba and $^{136}$Ba.
%Furthermore, in Fig. \ref{solar2} we show the percentage of $s$-process contributions also for the non $s$-only $^{135}$Ba, $^{137}$Ba, and $^{138}$Ba isotopes (blue dots).

%

% \begin{figure}[ht!]
% \plotone{Ba-isotopes-1.pdf}
% \caption{\label{solar2} Percentage of $s$-process contributions for both the $s$-only $^{134}$Ba and $^{136}$Ba isotopes (red dots) and the non $s$-only $^{135}$Ba, $^{137}$Ba, and $^{138}$Ba isotopes (blue dots), where we adopted the $\beta^-$-decay rates for $^{134}$Cs and $^{135}$Cs computed in the present work.}
% \end{figure} 

In conclusion, we find that the $\beta$-decay rates of the $^{134}$Cs and $^{135}$Cs isotopes remain lower (even by large factors for $^{134}$Cs) than the widely used TY estimates, thus pointing to longer half-lives at temperatures $T > 10^8$ K, which characterize the regions of the  AGB stars where the s-process nucleosynthesis occurs. This condition applies to both Cs isotopes even after considering that the rates are dramatically increased at high temperature by two concurrent factors: the inclusion of both the nuclear and the electronic ESs of parent and daughter nuclei, up to complete ionization. Most notably: i) the 60 keV nuclear ES of $^{134}$Cs, while scarcely populated, is the fastest to $\beta$-decay, with a rate $\simeq$80 times higher than the GS-to-GS one; ii) the rate increases with respect to GS decay only, close to a factor of $\simeq$3 at 20 keV, $\simeq$6 at 30 keV, $\simeq$8 at 40 keV.
For $^{135}$Cs similar effects, concurrent in lowering the half-life, were found, and our results are within the standard deviation of the general systematics. 

The use of the new rates to s-process computations in AGB stars remarkably reconciles models and observations for both the Sun and the presolar SiC grain isotopic composition by leading to an overall decrease in the $^{134}$Ba/$^{136}$Ba and $^{135}$Ba/$^{136}$Ba abundance ratios predictions, and provides realistic inputs and comparisons to the now forthcoming experiments \citep{PANDORA,mascali22}.
%Moreover, in addition to the temperature behavior of the Cs decay rates, we stress that at a given temperature our results differ significantly from those previously reported by TY. This increase with respect to TY is also responsible for the improvements in the agreement between observations and predictions of the AGB models about barium abundances.

\begin{acknowledgments}
DV acknowledges financial support from the German-Israeli Foundation (GIF No. I-1500-303.7/2019)
\end{acknowledgments}

\appendix
\section{Theoretical and computational model}\label{sec:app_theo}
Here we recap few mathematical and computational details of our method for calculating $\beta$-decay in astrophysical scenarios. In the traditional approach, $\beta$-decay spectra of allowed and forbidden transitions are calculated by using an analytical expression of the rate \citep{RevModPhys.90.015008} where several factors appear, which account for the nuclear and phase-space structure, and for the atomic exchange. Instead, our approach is based on the calculation of the total Hamiltonian of the system:
\begin{equation}
    \mathcal{H} = \mathcal{H}_\mathrm{nucl} + \mathcal{H}_\mathrm{e-e} + \mathcal{H}_\mathrm{weak},
    \label{eq:hamiltonian_general}
\end{equation}
where $\mathcal{H}_\mathrm{nucl}$ contains the interactions between nucleons in the initial and final nuclear states, 
$\mathcal{H}_\mathrm{e-e}$ is the electron-electron Coulomb correlation, taking into account that the transition occurs 
inside a partially ionized atom, and finally $\mathcal{H}_\mathrm{weak}$ is the
zero-order QED weak interaction, which fulfills the Lorentz-invariance:
\begin{equation} 
    \mathcal{H}_\text{weak} = \frac{G_F}{\sqrt{2}}H_\mu L^\mu + \mathrm{h.c.},
    \label{eq:weak_hamiltonian_cc}
\end{equation}
where $G_{F}=1.16637\times10^{-5}GeV^{-2}$, $L^\mu$ is the lepton current:
\begin{equation}
    L^\mu = \bar{u}_e \gamma^\mu(1-\gamma^5) v_\nu,
    \label{L_mu}
\end{equation}
and $H_\mu$ is the hadronic current:
\begin{equation}\label{hadr}
H_\mu=\bar{u}_p\gamma_{\mu}\left(1-x\gamma^{5}\right)v_{n}.
\end{equation}
In Eqs. (\ref{L_mu},\ref{hadr}) $\gamma^{\mu},\mu=0,...,4$ are the Dirac matrices, $x$ is the ratio between the axial vector coupling constant $c_{A}$ and the vector coupling constant $c_{V}$, $\frac{c_{A}}{c_{V}} \sim 1.2756$, and $u_e,v_\nu,v_n,u_p$ are the 
$\beta$-electron, neutrino, neutron and proton annihilation operators, respectively.

In particular, we are interested in reckoning the probability per unit time that the atomic system decays 
from a statistical mixture of initial states $\hat{\rho}_i=p_i|i><i|$ to a mixture of final states $P_f=\sum_{f}|f><f|$, 
that is: 
\begin{equation}\label{rate2}
N_{i\rightarrow f}=2\pi\mathrm{Tr}(\hat{\rho}_i \mathcal{H}_\text{weak} P_f\mathcal{H}_\text{weak})\delta(E_i-E_f)+h.c.
\end{equation}
In the typical approximation to the general theory of $\beta$-decay, where the recoil energy of the final nucleus is 
small with respect to the nucleon rest mass, the initial and final states can be written as:
\begin{eqnarray}
|i\rangle=|h_i\rangle\otimes |e_i\rangle\\
|f\rangle=|h_f\rangle\otimes |e_f\rangle\otimes |\bar{\nu}_f\rangle,
\end{eqnarray}
where $|h_{(i,f)}\rangle$ are initial and final multi-nucleon states characterized by the quantum numbers 
$|J_{(i,f)}, M_{(i,f)}, T_{(i,f)} \rangle_\mathrm{nucl}$,
where $J_{(i,f)},M_{(i,f)}$ denote the total angular momentum and its projection along some fixed axis of the initial 
and final states, respectively, and $T_{(i,f)}$ is the isospin; 
$|e_{i}\rangle=|j_{i}, m_{i}; \left[n_1^b \cdots n_k^b\right]_{(i)} \rangle_\mathrm{e-e}$ is the initial 
many-electron state, where $\left[n_1^b \cdots n_k^b\right]_{(i,f)}$ represent the bound orbitals; 
finally, $ |\bar{\nu}_f\rangle$ is the final anti-neutrino state (typically represented by a free plane wave).

We point out that the initial and final multi-nucleon states and the initial multi-electron states are characterized by a discrete spectrum, while the final multi-electron state, which describes the $\beta$ emission, is a continuum state that can be written as a linear combination of external products $ |e_f>=\sum_{j} I_{j,f} \wedge |\eta_{j,f}>$, where $|\eta_{j,f}>$ describes the $\beta$-electron continuum wavefunction and $I_{j,f}$ represents the final bound state.

Within this framework, the evaluation of the transition operator can be factorized into the product of lepton and 
hadronic currents \citep{morresi2018nuclear}. This is ultimately due to the large rest mass of the $W$ vector boson that 
mediates the weak interaction.
Furthermore, we can safely assume that electrons and neutrinos are not coupled and thus the lepton current (\ref{L_mu}) 
can be further factorized in the independent product of the neutrino and electron quantum field operators (or, dealing with 
expectation values, wavefunctions). In this work, we also make the assumption that the hadronic current can be factorized 
in the product of neutron and proton field operators. The hadronic current separability is basically equivalent to require 
that only one nucleon within the parent nucleus participates into the decay dynamics. The decaying nucleon is thus an 
independent particle, uncorrelated to the ``core'' of the remaining nucleons. Such a core, which couples only geometrically 
to the decaying nucleon so as to recover the total angular momentum of the parent reactant and of the final daughter nucleus, is approximated by a linear combination of angular momentum wavefunctions. Our  approximation assumes the validity of the nuclear shell model.
However, this does not represent an intrinsic limit of our method, as the hadronic current for systems where many-body effects 
are expected to play a paramount role can be assessed separately, via more accurate first-principle approaches \citep{HjorthJensen:391525}. 
The hadronic current (\ref{hadr}) is derived by an explicit numerical solution of the Dirac equation (DE) in a central potential. 
In our solution, protons and neutrons interact via a semi-empirical scalar and vector relativistic Wood-Saxon (WS) spherical symmetric potential, which describes the nucleon-nucleon interaction \citep{Woodsaxon}, whereby the DE is mono-dimensional in the radial variable. 
To solve the equation, a grid with a few thousands points is used \citep{morresi2018nuclear}.  

\begin{deluxetable}{@{\extracolsep{5pt}}ccccc}[t!]
\tablecaption{$n_e/n_p$ and $(n_e-n_p)/n_p$ ratios of proton densities $n_p=10^{26}$ and $n_p=10^{27}$ (cm$^{-3}$) and of the resulting electron densities $n_e$ as a function of temperatures $T$ (keV).\label{tab:tableS1}}
\tablehead{
\colhead{} & \multicolumn{2}{c}{$n_p=10^{26}$} & \multicolumn{2}{c}{$n_p=10^{27}$}  \\
\cline{2-3} \cline{4-5}
\colhead{$T$}&  $n_e/n_p$ & $(n_e-n_p)/n_p$ &  $n_e/n_p$ & $(n_e-n_p)/n_p$
}
\startdata
10   &  1.0   & 0.0 & 1.0 & 0.0 \\
20   &  1.0   &  2.121e-18 & 1.0 &     2.234e-21 \\
30   &  1.0  &  2.381e-10  &  1.0  &  2.57342e-13 \\
40   &  1.0  &  2.5098e-06 & 1.0  &   2.74123e-09  \\
50   & 1.0009 &  0.00087 & 1.0  &  9.5167e-07 \\
60   &  1.0460 &  0.0460 & 1.00005 &   5.3074e-05\\
70   &  1.5858  & 0.5858 & 1.001025 &  0.001025 \\
80   &  3.5663 &  2.5663 & 1.010026 &  0.010026 \\
90   &   8.0831  &  7.0831 & 1.0599 &  0.05987 \\
100  &  16.61334  & 15.61334 & 1.2334   & 0.2333676 \\
\enddata
\end{deluxetable}
Also the electron wavefunctions are found by solving self-consistently the Dirac-Hartree-Fock (DHF) equation in a central 
potential. Electrons interact via a mean-field, where the non-local exchange (Fock term) is replaced by the local density approximation 
(LDA) to the electron gas ($V_{ex}\propto \rho (r) ^{1/3} $) \citep{Slater1951,Salvat1987}. The numerical solution of the DHF equation was found by using a modified Runge-Kutta method \citep{morresi2018nuclear}.
The non-orthogonality between the bound initial $n_{i}^{b}$ and final $n_{f}^{b}$ orbitals, which are obtained by solving the DHF equations for the parent and daughter nuclei carrying a different atomic number, is taken into account. 
The continuum $\beta$-electron wavefunction $\psi_{f,e}^c(\bf{r})$ is then expressed in the field produced
by both the nucleus and the surrounding electrons as a Slater determinant, to take into account the atomic exchange.
The $\beta$-electron continuum wavefunction within this framework thus reflects the fact that the emitted lepton may decay into a 
bound state of the daughter nucleus with ejection of a secondary bound electron and that final state excitations, such as shake-up 
and shake-off \citep{morresi2018nuclear}, can be present.
Atomic exchange effects open up multiple decay channels that typically increase the $\beta$-decay rate particularly at low energies 
\citep{PhysRevA.45.6282}, where the overlap between the continuum and discrete wave functions maximizes.

Electrons populate the energy levels according to a Fermi-Dirac (FD) distribution 
$n_{e^-}^i=\frac{1}{1+e^{(\epsilon_i-\mu_{e^{-}})/(K_{\mathrm{B}}T)}}=F(T,\mu_{e^{-}})$, where $K_{\text B}$ is the 
Boltzmann constant and $T$ is the temperature. The eigenvalues $\epsilon_i$ are obtained by a self-consistent solution of the DHF 
equation for the leptons, and the chemical potential is assessed by assuming that the electrons behave as an ideal Fermi gas in a 
box with a relativistic energy-momentum dispersion 
$E^2=c^2p^2+m_e^2 c^4$ as $n_{e^-}=\int_0^{\infty}dp~ p^2/\pi^2\times(F((c\times \sqrt(p^2+m_e^2c^2)-\mu_e)/K_{\mathrm B}T)-F((c\times\sqrt(p^2+m_e^2c^2)+\mu_e)/K_{\mathrm{B}}T))$. 
We notice that the electron orbitals of Cs and Ba have not been re-optimized at each temperature.
At very high temperature, however, we included positron formation in our model.
The plasma at a given temperature is assumed overall neutral, that is $n_p=n_{e^-}-n_{e^+}$, where $n_p,n_{e^-},n_{e^+}$ are the proton, electron, and positron density, respectively. Note that $n_p \approx n_{e^-}$ at high densities, while they can differ sensibly a low densities.
Protons are treated as nonrelativistic particles and their density has been varied in the range $n_p=10^{24}-10^{27}$ to model different astrophysical scenarios. In Table \ref{tab:tableS1} we report the proton vs. relevant electron densities at different temperatures.
%\newpage

\section{Half-life tables}\label{sec:app_halflife}
Here we report the half-lives of both $^{134}$Cs and $^{135}$Cs on a finer mesh.
\begin{deluxetable}{ccccccc}[htb!]
\tablecaption{$^{134}$Cs half-lives (years) obtained by our model for several densities (cm$^{-3}$) and temperatures $T$ (keV).\label{tab:table4}}
\tablehead{
\colhead{$T$}& \multicolumn{6}{c}{$n_p$} \\
\cline{2-7}
\colhead{} & \colhead{\num{6e+24}} & \colhead{\num{6e+25}} & \colhead{\num{6e+26}} & \colhead{\num{6e+27}} & \colhead{\num{6e+28}} & \colhead{\num{6e+29}}
}
\startdata
  0.86   &   2.04149  & 2.05406  & 2.07755   & 2.15924  &  2.55484  & 5.56986 \\
   1.085  &   2.03336  & 2.05026 & 2.07621 & 2.15899 & 2.5548 & 5.56984 \\
   1.37 &  2.01435 & 2.04481 & 2.07446 & 2.15896 & 2.55518 & 5.57079 \\
   1.719 & 1.9947 &  2.03652 & 2.07328 &  2.16047 & 2.55763  & 5.57629 \\
   2.16 & 1.98896 &  2.02753& 2.07546 &  2.16711& 2.56659 & 5.59606 \\
   2.725 & 2.00225 &  2.02915  & 2.08667 &  2.18582& 2.59062 & 5.64877 \\
   3.43 &  2.03635 &      2.05478&     2.11409 &      2.2243 &     2.63937 &     5.75549 \\
   4.319 &  2.07298  &    2.10883 &   2.16471  &   2.2886   &   2.72088  &   5.93362 \\
  5.44 &   2.00961 &     2.17161&    2.23866 &   2.37687  &    2.83378 &    6.1752 \\
  6.845 &  1.86683 &     2.1586 &    2.30543 &   2.46085  &    2.94249 &   6.35983 \\
  8.62 &   1.78273 &   1.97719 &   2.24599 &  2.43095  &   2.90337 &  6.01333\\
  10.85 & 1.58026 & 1.64867 & 1.87921 & 2.08555  & 2.4593  & 4.55354 \\
  13.66 & 1.17738 & 1.19422 & 1.29305 & 1.45373  & 1.67893  & 2.70035 \\
  17.19 & 0.765335 & 0.769108 & 0.798542 & 0.886004  & 1.00964   & 1.47377 \\
  21.65 & 0.48386  & 0.484805 & 0.493239 & 0.533676  & 0.606594   & 0.84334 \\
  27.25 & 0.321219 & 0.321509& 0.324257& 0.342418 & 0.389374  & 0.530591 \\
  34.31 &  0.228839  & 0.228947 &    0.229995 &   0.238386  &  0.269924   &  0.365959 \\
 43.19 &   0.174786  & 0.174833 &      0.175296 &     0.179417 &    0.200815  &   0.27283 \\
  54.37 &   0.141613  &  0.141634 &      0.141865 &     0.144038 &    0.158478  &  0.215895 \\
  68.453  &   0.120307  & 0.120315 &      0.120432 &     0.121659 &    0.131326  &  0.178361 \\
  86.17 &  0.106109 & 0.106113&    0.106165&   0.106871&  0.113322&  0.151812 \\
\enddata
\end{deluxetable}

\begin{deluxetable}{ccccccc}[t!]
\tablecaption{$^{135}$Cs half-lives (years) obtained by our model for several densities (cm$^{-3}$) and temperatures $T$ (keV).\label{tab:table5}}
\tablehead{
\colhead{$T$}& \multicolumn{6}{c}{$n_p$} \\
\cline{2-7}
\colhead{} & \colhead{\num{6e+24}} & \colhead{\num{6e+25}} & \colhead{\num{6e+26}} & \colhead{\num{6e+27}} & \colhead{\num{6e+28}} & \colhead{\num{6e+29}}
}
\startdata
 0.86  &  2.24735e+06  &   2.28314e+06  &   2.35169e+06  &   2.5949e+06  &   4.05442e+06  &   1.38295e+08  \\
   1.085 &  2.22363e+06  &   2.27205e+06  &   2.34741e+06  &   2.59384e+06  &   4.05376e+06  &   1.38122e+08  \\
 1.37  &  2.16957e+06  &   2.25504e+06  &   2.3407e+06  &   2.59214e+06  &   4.05272e+06  &   1.37847e+08   \\
  1.719 &  2.11206e+06  &   2.22521e+06  &   2.33048e+06  &   2.58945e+06  &   4.05111e+06  &   1.37424e+0  \\
 2.16  &  2.07881e+06  &   2.17983e+06  &   2.31431e+06  &   2.58509e+06  &   4.04856e+06  &   1.36759e+08   \\
 2.725 &  2.06411e+06  &   2.1325e+06  &   2.28881e+06  &   2.57781e+06  &   4.04444e+06  &   1.35701e+08  \\
 3.43  &   2.0522e+06  &   2.09846e+06  &   2.25322e+06  &   2.56603e+06  &   4.03796e+06  &   1.34072e+08  \\
 4.319 &  1.97606e+06  &   2.07086e+06  &   2.21036e+06  &   2.54721e+06  &   4.02766e+06  &   1.3157e+08  \\
  5.44 &  1.62748e+06  &   1.9987e+06  &   2.16292e+06  &   2.51833e+06  &   4.01129e+06  &   1.27794e+08  \\
  6.845 &  1.24302e+06  &   1.75501e+06  &   2.09403e+06  &   2.47574e+06  &   3.9854e+06  &   1.22285e+08  \\
8.62  &  1.10909e+06  &   1.39724e+06  &   1.94529e+06  &   2.41137e+06  &   3.94404e+06  &   1.14477e+08  \\
 10.85 &  1.07248e+06  &   1.18395e+06  &   1.67416e+06  &   2.29764e+06  &   3.85507e+06  &   9.81098e+07  \\
 13.66 &  806399  &   834310  &   1.02279e+06  &   1.44121e+06  &   2.24393e+06  &   1.27362e+07  \\
17.19 &  72041.5  &   72659.3  &   77637.2  &   94275.4  &   123468  &   340712  \\
 21.65 &  3851.17  &   3865.52  &   3995.78  &   4683.27  &   6254.44  &   16801.2  \\
 27.25 &  358.829  &   359.506  &   365.995  &   412.098  &   563.022  &   1513.15  \\
 34.31 &  53.6033  &   53.6588  &   54.2028  &   58.7768  &   80.1084  &   214.686  \\
 43.19 &  11.6125  &   11.6195  &   11.6889  &   12.3265  &   16.2948  &   42.9303  \\
 54.37 &  3.34727  &   3.34841  &   3.36067  &   3.47855  &   4.37922  &   11.0515  \\
 68.453 &  1.20894  &   1.20911  &   1.2117  &   1.23923  &   1.47913  &   3.46087  \\
  86.17 &  0.527193  &   0.527237  &   0.527765  &   0.535284  &   0.609228  &   1.28288  \\
\enddata
\end{deluxetable}

\begin{deluxetable}{ccc}[t!]
\tablecaption{$^{134}$Cs and $^{135}$Cs half-lives (years) obtained by our model at Earth conditions without (wcf) the correction factor in comparison to experimental data (Obs.). The correction factors to reproduce the measured half-lives are 4.4 and $\approx$90 for $^{134}$Cs and $^{135}$Cs, respectively. \label{tab:table6}}
\tablehead{
\colhead{} & \colhead{Obs.} & \colhead{wcf} 
}
\startdata
 $^{134}$Cs  &   2.0652  & 8.80   \\
 $^{135}$Cs  &  2.3e+06   & 0.025e+06   \\
\enddata
\end{deluxetable}

\FloatBarrier

\bibliography{biblio}{}
\bibliographystyle{aasjournal}

%% This command is needed to show the entire author+affiliation list when
%% the collaboration and author truncation commands are used.  It has to
%% go at the end of the manuscript.
%\allauthors

%% Include this line if you are using the \added, \replaced, \deleted
%% commands to see a summary list of all changes at the end of the article.
%\listofchanges

\end{document}